\documentclass[a4paper]{jpconf}
\usepackage{graphicx}
\usepackage{amssymb}
\usepackage{amsmath}
\usepackage{tabularx}

\def\AEF{A.E. Faraggi}
\def\JHEP#1#2#3{{\it JHEP}\/ {\bf #1} (#2) #3}
\def\NPB#1#2#3{{\it Nucl.\ Phys.}\/ {\bf B#1} (#2) #3}
\def\PLB#1#2#3{{\it Phys.\ Lett.}\/ {\bf B#1} (#2) #3}
\def\PRD#1#2#3{{\it Phys.\ Rev.}\/ {\bf D#1} (#2) #3}
\def\PRL#1#2#3{{\it Phys.\ Rev.\ Lett.}\/ {\bf #1} (#2) #3}

\def\IJMP#1#2#3{{\it Int.\ J.\ Mod.\ Phys.}\/ {\bf A#1} (#2) #3}

\def\EJP#1#2#3{{\it Eur.\ Phys.\ Jour.}\/ {\bf C#1} (#2) #3}

\begin{document}
\rightline{LTH--1038}
\rightline{March 2015}
	
\title{Exotica and discreteness in the classification of string spectra}

\author{Hasan Sonmez}

\address{Department of Mathematical Sciences, University of Liverpool, Liverpool L69 7ZL, United Kingdom}

\ead{Hasan.Sonmez@Liverpool.ac.uk}

\begin{abstract}
I discuss the existence of discrete properties in the landscape of free fermionic heterotic-string vacua that were discovered via their classification by $SO(10)$ GUT models and its subgroups such as the Pati-Salam, Flipped $SU(5)$ and $SU(4) \times SU(2) \times U(1)$ models. The classification is carried out by fixing a set of basis vectors and varying the GGSO projection coefficients entering the one-loop partition function. The analysis of the models is facilitated by deriving algebraic expressions for the GSO projections that enable a computerised analysis of the entire string spectrum and the scanning of large spaces of vacua. The analysis reveals discrete symmetries like the spinor-vector duality observed at the $SO(10)$ level and the existence of exophobic Pati-Salam vacua. Contrary to the Pati-Salam case the classification shows that there are no exophobic Flipped $SU(5)$ vacua with an odd number of generations. It is observed that the $SU(4) \times SU(2) \times U(1)$ models are substantially more constrained.
\end{abstract}

\section{Introduction}

The four space-time dimensional free fermionic construction \cite{fff} of the heterotic-string provides a worldsheet approach to analysing quasi-string vacua. The models constructed to date corresponding to symmetric and asymmetric ${\mathbb{Z}}_2 \times {\mathbb{Z}}_2$ orbifold compactifications represent some of the most realistic string models consisting of three generations. The early quasi-examples were built since the late eighties, that were composed of asymmetric ${\mathbb{Z}}_2 \times {\mathbb{Z}}_2$ orbifold compactifications. These models corresponded to compactifications with ${\cal N}=(2,0)$ super-conformal worldsheet symmetry, where the observable symmetry was taken to be a $E_8$ gauge group and then broken down to a $SO(10)$ symmetry subgroup. The cases consisted of $SU(5)\times U(1)$ (Flipped $SU(5)$) \cite{revamp,fsu5}, $SO(6)\times SO(4)$ (Pati-Salam) \cite{alr}, $SU(3)\times SU(2)\times U(1)^2$ (Standard-Like) \cite{slm}, and $SU(3)\times SU(2)^2\times U(1)$ (Left-Right symmetric) \cite{lrs}. Some of these models shared the NAHE-based structure \cite{nahe}, which was used to develop the contemporary research in the free fermionic model building that focuses on exploring large classes of string vacua. In the end of the nineties tools for the classification of the free fermionic symmetric ${\mathbb{Z}}_2 \times {\mathbb{Z}}_2$ orbifolds were derived for type II superstrings \cite{gkr}
and extended similarly in the heterotic-string construction \cite{fknr,fkr} during the last decade.

The classification of the heterotic-string vacua with unbroken $E_6$ and $SO(10)$ GUT gauge groups, revealed the existence of a symmetry called the spinor-vector duality in the space of ${\mathbb{Z}}_2 \times {\mathbb{Z}}_2$ (and ${\mathbb{Z}}_2$) string models. That is under the $SO(10)$ the exchange of spinorial $\bf{16}$ plus anti-spinorial $\bf{\overline{16}}$ and vectorial $\bf{10}$ representations \cite{fkr,cfkr,xmap,rizos}. Then the classification was extended, where the $SO(10)$ symmetry was broken to the Pati-Salam subgroup in \cite{acfkr}. This revealed that exophobic string vacua existed for all the sectors containing the massless states and the exotic fractionally charged fermions appeared only in the massive spectrum. An example of a three generation Pati-Salam model was studied in \cite{cfr} and was shown to be phenomenologically viable. This classification method similar to the Pati-Salam models \cite{cfr}, was used to study $E_6$ symmetries broken down the maximal $SU(6)\times SU(2)$ subgroup \cite{su62}. Here an exophobic model was found that admitted an additional anomaly free family universal $U(1)$ symmetry beyond the $U(1)$ generators of the $SO(10)$ GUT gauge group \cite{anomfreeu1}. However, further studies involving the classifications of the Flipped $SU(5)$ \cite{frs} and $SU(4) \times SU(2) \times U(1)$ \cite{fs} subgroups of $SO(10)$ was shown to contain no exophobic string vacua. The Flipped $SU(5)$ models was shown to only produce exophobic models with even generations and only three generation models with exophilic vacua. The $SU(4) \times SU(2) \times U(1)$ models produces no three generation models as all right-handed particles were projected out of the massless spectrum \footnote{Similarly in \cite{cfn} it was shown that for the $SU(4) \times SU(2) \times U(1)$ gauge group, three generation models are forbidden in NAHE based basis vectors}.

The methodologies developed in the classifications of \cite{gkr,fknr,fkr,rizos,acfkr,frs,fs} provided a vital tool to analyse the phenomenological properties of large classes of string vacua \footnote{Other groups have also performed analysis of large sets of string vacua \cite{statstud}}. In this paper, I discuss the techniques used to classify these large classes of string vacua and the landscape of the free fermionic models in which the $SO(10)$ symmetry is broken to its subgroups. I will firstly construct $SO(10)$ models with the required basis vectors consistent with the ABK rules \cite{fff} containing only period and anti-periodic fermion boundary conditions. Then add the additional $SO(10)$ symmetry breaking basis vectors to form the subgroups: Pati-Salam, Flipped $SU(5)$, $SU(4) \times SU(2) \times U(1)$ and Standard-Like models. This will follow by detailed properties of the Flipped $SU(5)$ classification and the discussion of the classification results for the existence of three generation models in the free fermionic landscape.

\section{Free Fermionic Models}\label{analysis}

The free fermionic models corresponding to ${\cal N}=(2,0)$ super-conformal worldsheet symmetry in the ${\mathbb{Z}}_2 \times {\mathbb{Z}}_2$ orbifold compactifications is fixed at an extended
symmetry point in the moduli space. Here the directions that are compactified is represented by the fermions propagating in the two dimensional string worldsheet \cite{z2z2one,z2z2two}.
The free fermionic construction provides an elegant approach to studying phenomenologically viable properties of the string vacua. The matter content arises from the $\bf{27}$ of the $E_6$ symmetry, which breaks to the $SO(10)$ symmetry at the string scale decomposing under the $\bf{16}$ spinorial and $\bf{10}$ vectorial representations. Here the $\bf{16}$ consists of all the left- and right-handed fermions known (and predicted) at low energies and the $\bf{10}$ contains the Higgs states. The $SO(10)$ gauge group is further broken at the string scale to one of its subgroup and therefore the gauge group in the effective low energy field theory is given to be a subgroup of $SO(10)$.

\subsection{\bf{Free Fermionic Construction}}
The four-dimensional free fermionic construction in the light-cone gauge is represented by $20$ left-moving and $44$ right-moving real worldsheet fermions. These worldsheet fermions acquire a phase as they are parallel transported along the non-contractible loops of the vacuum to vacuum amplitude. The usual light-cone gauge notation of the fermions are given by:
$\psi^\mu_{1,2}, \chi^{1,\dots,6},y^{1,\dots,6}, \omega^{1,\dots,6}$
(left-movers) and
$\overline{y}^{1,\dots,6},\overline{\omega}^{1,\dots,6}$,
$\psi^{1,\dots,5}$, $\overline{\eta}^{1,2,3}$, $\overline{\phi}^{1,\dots,8}$
(right-movers). Under the modular invariance constraints \cite{fff}, each model is defined by a particular choice of phases for the fermions that can be spanned by a set of basis vectors
$v_{1},\dots,v_{N}$
$$v_i=\left\{\alpha_i(f_1),\dots,\alpha_i(f_{20})|\alpha_i(\overline{f}_1),
\dots,\alpha_i({\overline{f}_{44}})\right\}.$$
The basis vectors generate a space $\Xi$ which produces the string spectrum consisting of $2^{N}$ sectors. Each sector here is given as a linear combination of all the basis vectors 
\begin{equation}
\xi = \sum_{i=1}^N m_j v_i, \,\,\,\,\,\,\,\,\,\, m_j = 0,1,\dots,N_j-1,
\label{Xi}
\end{equation}
where $N_j \cdot v_j = 0$ mod $2$. They also describe the transformation properties of each fermion on the worldsheet, which is given by
\begin{equation}
	f_j\to -e^{i\pi\alpha_i(f_j)}\ f_j,  \,\,\,\,\,\,\,\,\,\, j = 1,\dots,64.
\end{equation}
The basis vectors also induces the generalised GSO projections,
with an action on any given string state $|S_{\xi}>$. This can be written as
\begin{equation}\label{gso}
	e^{i\pi v_i\cdot F_{\xi}} |S_{\xi}> = \delta_{{\xi}}\ C \binom {\xi} {v_i}^* |S_{\xi}>
\end{equation}
where $\delta_{{\xi}}=\pm1$
is the index for the space-time spin statistics and $F_{\xi}$ is the fermion number operator. Varying the different set of GGSO projection coefficients $ C \binom {\xi}{v_i}=\pm1$ produces distinct string models. In summary, a free fermionic model is constructed by a set of basis vectors $v_{1},\dots,v_{N}$ together with a set of $2^{N(N-1)/2}$ independent GGSO projection coefficients $C \binom{v_i}{v_j}, i>j$ consistent with modular invariance.

\subsection{\bf{SO(10) Models}}

The $SO(10)$ GUT models in the symmetric ${\mathbb{Z}}_2 \times {\mathbb{Z}}_2$ orbifold free fermionic construction are generated by a set of 12 basis vectors. The $SO(10)$ symmetry is then given by the following set of basis vectors

\begin{eqnarray}\label{SO10basis}
v_1={\bf1}&=&\{\psi^\mu,\
\chi^{1,\dots,6},y^{1,\dots,6}, \omega^{1,\dots,6}| \nonumber\\
& & ~~~\overline{y}^{1,\dots,6},\overline{\omega}^{1,\dots,6},
\overline{\eta}^{1,2,3},
\overline{\psi}^{1,\dots,5},\overline{\phi}^{1,\dots,8}\},\nonumber\\
v_2=S&=&\{{\psi^\mu},\chi^{1,\dots,6}\},\nonumber\\
v_{2+i}={e_i}&=&\{y^{i},\omega^{i}|\overline{y}^i,\overline{\omega}^i\}, \
i=1,\dots,6,\nonumber\\
v_{9}={b_1}&=&\{\chi^{34},\chi^{56},y^{34},y^{56}|\overline{y}^{34},
\overline{y}^{56},\overline{\eta}^1,\overline{\psi}^{1,\dots,5}\},\label{basis}\\
v_{10}={b_2}&=&\{\chi^{12},\chi^{56},y^{12},y^{56}|\overline{y}^{12},
\overline{y}^{56},\overline{\eta}^2,\overline{\psi}^{1,\dots,5}\},\nonumber\\
v_{11}=z_1&=&\{\overline{\phi}^{1,\dots,4}\},\nonumber\\
v_{12}=z_2&=&\{\overline{\phi}^{5,\dots,8}\}.
\nonumber
\end{eqnarray}
Similarly the asymmetric ${\mathbb{Z}}_2 \times {\mathbb{Z}}_2$ orbifolds can be constructed \footnote{Asymmetric ${\mathbb{Z}}_2 \times {\mathbb{Z}}_2$ orbifold basis vectors can be found in \cite{revamp,fsu5,alr,slm,lrs,nahe} for the different GUT models}. The first basis vector ${\bf1}$ in equations (\ref{SO10basis}) is a requirement of the ABK rules \cite{fff} in order to preserve modular invariance; this generates the $SO(44)$ gauge symmetry together with tachyons in the massless string spectrum. With the addition of vector $S$, the tachyons are all projected out the massless spectrum as we construct a ${N} = 4$ supersymmetric theory as well as preserving the $SO(44)$ gauge group. The following six vectors: $e_{1}$,$\dots$,$e_{6}$ all correspond to the possible symmetric shifts of the six internal coordinates, this therefore breaks the $SO(44)$ group to $SO(32) \times U(1)^6$ but preserves ${N = 4}$ space-time supersymmetry. The vectors $b_1$ and $b_2$ on the ${\mathbb{Z}}_2 \times {\mathbb{Z}}_2$ orbifold corresponds to the twists, which break ${N} = 4$ to ${N} = 1$ space-time supersymmetry. These vectors also break the $U(1)^6$ symmetry giving rise to the $SO(10) \times U(1)^2 \times SO(18)$ gauge symmetry. The states coming from the hidden sector are produced by the vectors $z_1$ and $z_2$, which is given by the remaining fermions: $\overline{\phi}^{1,\dots,8}$, that were not affected by the action of the previous vectors on the GGSO projection given in equation (\ref{gso}). These vectors together with the others generate the following adjoint representation of the gauge symmetry: $SO(10) \times U(1)^3 \times SO(8) \times SO(8)$ where $SO(10) \times U(1)^3$ is the observable gauge group that gives rise to the matter states arising from the twisted sectors, that is charged under the $U(1)$s and $SO(8) \times SO(8)$ the hidden gauge group where all the matter states are neutral under the $U(1)$s.

\subsection{\bf{SO(10) Subgroups}}

The $SO(10)$ GUT models generated by equation (\ref{basis}) can be broken 
to its subgroup by the boundary condition assignment of the complex 
fermions $\overline{\psi}^{1,\dots,5}$. For the Flipped $SU(5)$ and Pati-Salam cases we require one additional basis vector that we denote as $v_N\equiv\alpha$ to break the $SO(10)$ symmetry.
In these cases, the $SO(6)\times SO(4)$ models which were classified in \cite{acfkr} utilised solely periodic and anti-periodic boundary conditions, whereas in the Flipped $SU(5)$ case that were classified in \cite{frs}, the boundary conditions included ${1}/{2}$ assignments. In order to construct the $SU(4) \times SU(2) \times U(1)$ and Standard-Like models, we require the Pati-Salam breaking as well as an additional $SO(10)$ breaking basis vector. These gauge groups can be constructed from the following basis vectors: \newline
\indent {\underline{Pati-Salam Models:}}
\begin{equation} \label{patisalambasis}
v_{13} \,\,\, =\alpha \, = \{ \overline{\psi}^{4,5},\overline{\phi}^{1,2} \}.
\end{equation}
\indent {\underline{Flipped SU(5) Models:}}
\begin{equation} \label{su5basis}
v_{13} \,\,\, =\alpha \, = \{ \overline{\eta}^{1,2,3} = \frac{1}{2},\overline{\psi}^{1,...,5} = \frac{1}{2},\overline{\phi}^{1,...,4} = \frac{1}{2},\overline{\phi}^{5} = 1\}.
\end{equation}
\indent {\underline{SU(4) x SU(2) x U(1) Models:}}
\begin{eqnarray} \label{su421basis}
v_{13} \,\,\, &=&\alpha \, = \{ \overline{\psi}^{4,5},\overline{\phi}^{1,2} \}, \nonumber\\
v_{14} \,\,\, &=&\beta \, = \{ \overline{\psi}^{4,5} = \frac{1}{2},\overline{\phi}^{1,...,6} = \frac{1}{2}\}.
\end{eqnarray}
\indent {\underline{Standard-Like Models:}}
\begin{eqnarray} \label{su3211basis}
v_{13} \,\,\, &=&\alpha \, = \{ \overline{\psi}^{4,5},\overline{\phi}^{1,2} \},  \nonumber\\
v_{14} \,\,\, &=&\beta \, = \{ \overline{\eta}^{1,2,3} = \frac{1}{2},\overline{\psi}^{1,...,5} = \frac{1}{2},\overline{\phi}^{1,...,4} = \frac{1}{2},\overline{\phi}^{5} = 1\}.
\end{eqnarray}
Hereafter, we will solely consider the study of the Flipped $SU(5)$ models and present the classification methodology using this $SO(10)$ subgroup.

\section{Classification Methodology}\label{classification}
The classification methodology facilitates a straight forward scanning of large amounts of string vacua, where all the twisted sectors are originally identified and then checked one by one for viable phenomenological properties. These sectors contain the observable, hidden and exotic sectors.
The hidden sectors consists of only singlets under the $SU(5)$ symmetry and are neutral under all the observable $U(1)$ gauge groups, therefore all hidden matter content is charged under a $U(1)$ hidden gauge group only. The exotic sectors on the other hand consist of fractionally charged massless states \cite{ww,schellekens,fc} under $SU(5) \times U(1)$ gauge group and thus, these states do not fall into representations that are predicted by the Standard Model \cite{halyo} and are therefore excluded from the free fermionic string vacua if possible. 

\subsection{\bf{Observable Matter Spectrum}}
The observable sectors give rise to the chiral matter content, where the $SO(10)$ $\bf{16}$ and $\overline{\bf{16}}$ spinorial representations are given by the following 48 sectors:
\begin{eqnarray} \label{obspin}
B_{pqrs}^{(1)}&=& S + {b_1 + p e_3+ q e_4 + r e_5 + s e_6} \nonumber\\
&=&\{\psi^\mu,\chi^{12},(1-p)y^{3}\overline{y}^3,p\omega^{3}\overline{\omega}^3,
(1-q)y^{4}\overline{y}^4,q\omega^{4}\overline{\omega}^4, \nonumber\\
& & ~~~(1-r)y^{5}\overline{y}^5,r\omega^{5}\overline{\omega}^5,
(1-s)y^{6}\overline{y}^6,s\omega^{6}\overline{\omega}^6,
\overline{\eta}^1,\overline{\psi}^{1,..,5}\},
\\
B_{pqrs}^{(2)}&=& S + {b_2 + p e_1+ q e_2 + r e_5 + s e_6},
\label{twochiralspinorials}
\nonumber\\
B_{pqrs}^{(3)}&=& S + {b_3 + p e_1+ q e_2 + r e_3 + s e_4}. \nonumber
\end{eqnarray}
where $p,q,r,s=0,1$ and $b_3=b_1+b_2+2\alpha+z_1$. These 48 sectors are broken such that the $\textbf{16}$ and $\overline{\textbf{16}}$ spinorial representations of $SO(10)$ decomposed under $SU(5) \times U(1)$ are given by 
\begin{eqnarray}
\textbf{16} &= &\left(\overline{\textbf{5}},
-{\tfrac{{3}}{{2}}}\right) + 
\left(\textbf{10},+{\tfrac{{1}}{{2}}}\right) + 
\left(\textbf{1},+{\tfrac{{5}}{{2}}}\right), \nonumber \\
\overline{\textbf{16}} &= &\left(\textbf{5},
+{\tfrac{{3}}{{2}}}\right) + 
\left(\overline{\textbf{10}},-{\tfrac{1}{2}}\right)
+ \left(\textbf{1},-{\tfrac{{5}}{{2}}}\right).
\end{eqnarray}
In the free fermionic Flipped $SU(5)$ models, the hypercharge and the electromagnetic charge is given by the following normalisations:
\begin{eqnarray}
Y &=& \frac{1}{3} (Q_1 + Q_2 + Q_3) + \frac{1}{2} (Q_4 + Q_5), \nonumber\\
Q_{em} &=& Y + \frac{1}{2} (Q_4 - Q_5),
\end{eqnarray}
where the charges $Q_{i}$ from a given string state, arise due to the $\overline{\psi}^{i}$ complex fermions for $i = 1,...,5$. The charges of the colour $SU(3)$ and electroweak $SU(2) \times U(1)$ Cartan generators, of the matter content in Flipped $SU(5) \times
U(1)$ symmetry is summarised in the following table:

\begin{center}
	\begin{tabular}{|c|c|c|c|c|}
		\hline
		Representation & $\overline{\psi}^{1,2,3}$ &
		$\overline{\psi}^{4,5}$ & $Y$ & $Q_{em}$ \\
		\hline \hline
		& $(+,+,+)$ & ($+,-$)& 1/2& 1,0\\
		$\left( \, \textbf{5} \, , \, +\frac{{3}}{{2}} \, \right)$ & ($+,+,-$)& $(+,+)$ & 2/3& 2/3\\ 
		\hline
		& ($+,-,-$)& $(-,-)$ & -2/3& -2/3\\
		$\left( \, \overline{\textbf{5}} \, , \, -\frac{{3}}{{2}} \, \right)$ & $(-,-,-)$ & ($+,-$)& -1/2& -1,0\\ 
		\hline
		& $(+,+,+)$ & $(-,-)$ & 0& 0\\
		$\left(\textbf{10},+\frac{{1}}{{2}}\right)$& ($+,-,-$)& $(+,+)$ & 1/3& 1/3\\
		& ($+,+,-$)& ($+,-$)& 1/6& -1/3,2/3\\ 
		\hline
		& ($+,+,-$)&
		$(-,-)$ & -1/3& -1/3\\
		$\left(\overline{\textbf{10}},-\frac{{1}}{{2}}\right)$ & ($+,-,-$)& ($+,-$)& -1/6& 1/3,-2/3\\
		& $(-,-,-)$ & $(+,+)$ & 0& 0\\ 
		\hline
		$( \, \textbf{1} \, , \, +\frac{{5}}{{2}} \, )$&
		$(+,+,+)$ & $(+,+)$ & 1& 1\\ 
		\hline
		$( \, \textbf{1} \, , \, -\frac{{5}}{{2}} \, )$
		& $(-,-,-)$ & $(-,-)$ & -1& -1\\ 
		\hline
	\end{tabular}
\end{center}
Using this table we can deduce that the states corresponding to the particles of the Standard Model, are decomposed under the $SU(3) \times SU(2) \times U(1)$ gauge group are as follows:
\begin{align} \label{16decomposition}
\left( \,  \,  \overline{\textbf{5}} \,  ,-\frac{3}{2}\right)&
=\left(\overline{\textbf{3}},\textbf{1},-\frac{2}{3}\right)_{u^c}+\left(\textbf{1},\textbf{2},-\frac{1}{2}\right)_{L}, \nonumber \\
\left(\textbf{10},+\frac{1}{2}\right)&=\left(\textbf{3},\textbf{2},+\frac{1}{6}\right)_{Q} \, 
+\left(\overline{\textbf{3}},\textbf{1},+\frac{1}{3}\right)_{d^c}+\left(\textbf{1},\textbf{1},0\right)_{\nu^c},\\
\left( \,  \,  \textbf{1} \, ,+\frac{5}{2}\right)&=\left(\textbf{1},\textbf{1},+1 \, \right)_{e^c}, \nonumber
\end{align}
where the subscripts: $Q$ and $L$ are the left-handed quark- and lepton-doublets respectively; $d^c,~u^c,~e^c$ and $\nu^c$ are the right-handed quark and lepton singlets.

\subsection{\bf{Gauge Group Enhancements}}\label{gge}
The untwisted sectors in the string spectrum give rise to the space-time vector bosons. In the Flipped $SU(5)$ models the untwisted Neveu schwarz (NS) sector generates the following gauge groups
$$
SU(5)\times U(1)\times{U(1)}^3\times SU(4) \times U(1) \times U(1) \times SO(6).
$$
Depending on the GGSO projection coefficients chosen, the following twelve untwisted sectors can enhance the above gauge group to a higher gauge group:
\begin{equation}\label{ggsectors}
	\mathbf{G} =
	\left\{ \begin{array}{ccccc}
		\,\,\,\, z_1          ,&
		\,\,\,\, z_2          ,&
		\,\,\,\, z_1 + z_2    ,&
		\,\,\,\, z_1 + 2\alpha,\\
		\,\,\,\,\, \alpha       ,&
		\,\,\,\,\, z_1 + \alpha ,&
		\,\,\, z_2 + \alpha ,&
		\,\, z_1 + z_2 + \alpha,\\
		\,\, 3\alpha       ,&
		\,\,\,\,\,\,\,\, z_1 + 3\alpha ,&
		\,\,\,\,\,\, z_2 + 3\alpha ,&
		\,\,\, z_1 + z_2 + 3\alpha
	\end{array} \right\}. 
\end{equation}
We impose the GGSO projection coefficients such that the above sectors in equation (\ref{ggsectors}) are all projected out. Therefore the string vacua are free of gauge group enhancements and thus the Flipped $SU(5)$ symmetry is preserved. For example the sector $z_1 + 2 \alpha = \{ \overline{\psi}^{1,...,5}, \overline{\eta}^{1,2,3} \}$ contributes to the enhancement of the observable gauge group consisting of $SU(5) \times U(1)^4$. The following table shows the conditions avoided and thus eliminating the resulting gauge group enhancement
\begin{center}
	\begin{tabularx}{\textwidth}{|X|}
		
		\hline
		\textbf{Sector Condition}  \\ \hline
		$(z_1 + 2 \alpha|e_i) = (z_1 + 2 \alpha|z_k) = 0$
		\\ \hline
	\end{tabularx}
	
	\begin{tabularx}{\textwidth}{|X|X|}
		
		\hline
		\textbf{Enhancement Condition} &  \textbf{Resulting Enhancement} \\ \hline
		$(z_1 + 2 \alpha|\alpha) \, = (z_1 + 2 \alpha|b_2)$ & 
		$SU(5)_{obs}\times U(1)_5 \times U(1)_{\zeta} $
		\newline
		$ \xrightarrow{\hspace*{1cm}} SU(6) \times SU(2) $ \\ \hline
		$(z_1 + 2 \alpha|\alpha) \, \neq (z_1 + 2 \alpha|b_2)$ & 
		$SU(5)_{obs}\times U(1)_5 \times U(1)_{\zeta} \newline \xrightarrow{\hspace*{1cm}} SO(10) \times U(1)$ \\ \hline
	\end{tabularx}
\end{center}
where $i= 1,\dots,6$, $k=1,2$ and $U(1)_{\zeta}$ is a linear combination of $U(1)_{1}$, $U(1)_{2}$ and $U(1)_{3}$.

\subsection{\bf{Projectors}}

The states in the sector $B_{pqrs}^{(1,2,3)}$ as given in equation (\ref{obspin}), can be projected in or out of the string spectrum easily, by the GGSO projection coefficients on the basis vectors $e_1$, $e_2$, $z_1$ and $z_2$. Hence we define a projector called $P$, such that if the states survive, $P=1$ and if the states are projected out, $P=0$:

\footnotesize
\begin{align}
P_{pqrs}^{(1)} &= \frac{1}{16} 
\left( 1-C \binom {e_1} {B_{pqrs}^{(1)}}\right) . 
\left( 1-C \binom {e_2} {B_{pqrs}^{(1)}}\right) . 
\left( 1-C \binom {z_1} {B_{pqrs}^{(1)}}\right) . 
\left( 1-C \binom {z_2} {B_{pqrs}^{(1)}}\right), \nonumber\\
P_{pqrs}^{(2)} &= \frac{1}{16} 
\left( 1-C \binom {e_3} {B_{pqrs}^{(2)}}\right) . 
\left( 1-C \binom {e_4} {B_{pqrs}^{(2)}}\right) . 
\left( 1-C \binom {z_1} {B_{pqrs}^{(2)}}\right) . 
\left( 1-C \binom {z_2} {B_{pqrs}^{(2)}}\right),\\
P_{pqrs}^{(3)} &= \frac{1}{16} 
\left( 1-C \binom {e_5} {B_{pqrs}^{(3)}}\right) . 
\left( 1-C \binom {e_6} {B_{pqrs}^{(3)}}\right) . 
\left( 1-C \binom {z_1} {B_{pqrs}^{(3)}}\right) . 
\left( 1-C \binom {z_2} {B_{pqrs}^{(3)}}\right). \nonumber
\end{align}
\normalsize
Further derivation shows that these projectors can be rewritten as a system of linear equations with $p$, $q$, $r$ and $s$ as unknowns, which can be expressed in the following way:

\begin{align} \label{matrix}
\begin{pmatrix} (e_1|e_3)&(e_1|e_4)&(e_1|e_5)&(e_1|e_6)\\
(e_2|e_3)&(e_2|e_4)&(e_2|e_5)&(e_2|e_6)\\
(z_1|e_3)&(z_1|e_4)&(z_1|e_5)&(z_1|e_6)\\
(z_2|e_3)&(z_2|e_4)&(z_2|e_5)&(z_2|e_6) \end{pmatrix}
\begin{pmatrix} p\\q\\r\\s\end{pmatrix} &=
\begin{pmatrix} (e_1|b_1)\\
(e_2|b_1)\\
(z_1|b_1)\\
(z_2|b_1)
\end{pmatrix}, \nonumber
\\[0.3cm]
\begin{pmatrix} (e_3|e_1)&(e_3|e_2)&(e_3|e_5)&(e_3|e_6)\\
(e_4|e_1)&(e_4|e_2)&(e_4|e_5)&(e_4|e_6)\\
(z_1|e_1)&(z_1|e_2)&(z_1|e_5)&(z_1|e_6)\\
(z_2|e_1)&(z_2|e_2)&(z_2|e_5)&(z_2|e_6) \end{pmatrix}
\begin{pmatrix} p\\q\\r\\s\end{pmatrix} &=
\begin{pmatrix} (e_3|b_2)\\
(e_4|b_2)\\
(z_1|b_2)\\
(z_2|b_2)
\end{pmatrix},
\\[0.3cm]
\begin{pmatrix} (e_5|e_1)&(e_5|e_2)&(e_5|e_3)&(e_5|e_4)\\
(e_6|e_1)&(e_6|e_2)&(e_6|e_3)&(e_6|e_4)\\
(z_1|e_1)&(z_1|e_2)&(z_1|e_3)&(z_1|e_4)\\
(z_2|e_1)&(z_2|e_2)&(z_2|e_3)&(z_2|e_4) \end{pmatrix}
\begin{pmatrix} p\\q\\r\\s\end{pmatrix} &=
\begin{pmatrix} (e_5|b_3)\\
(e_6|b_3)\\
(z_1|b_3)\\
(z_2|b_3)
\end{pmatrix}. \nonumber
\end{align}
Here we defined the GGSO phases as
$$C \binom{v_i}{v_j} = e^{i \pi (v_i|v_j)}~$$
where $v_i$ and $v_j$ are basis vectors. If we write the above equation in (\ref{matrix}) as $\Delta^{i}W^{i} = Y^{i}$ for $i=1,2,3$, then using linear algebra methods we can simply check if Rank(Matrix[$\Delta^{i}$])$=$Rank(AugmentedMatrix[$\Delta^{i},Y^{i}$]) and when it's equal, then we have $2^{4-\mbox{Rank}(\mbox{Matrix}[\Delta^{i}])}$ solutions to the equation.

\section{\bf{Classification Results}}
\begin{figure}[!]
	\centering
	\includegraphics[width=140mm]{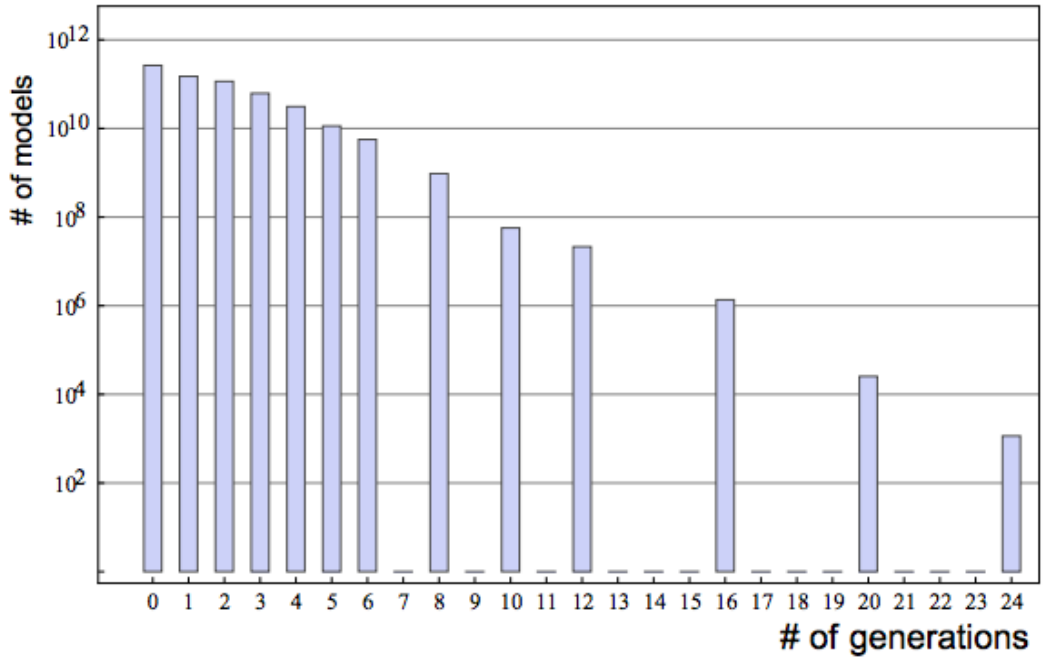}
	\caption{\emph{Number of exophobic models versus number of generations in a random sample of $10^{11}$ $SO(10)$ configurations as given in Figure 1 in \cite{rizos}.}}
	\label{figure1}
\end{figure}

\begin{figure}[!]
	\centering
	\includegraphics[width=140mm]{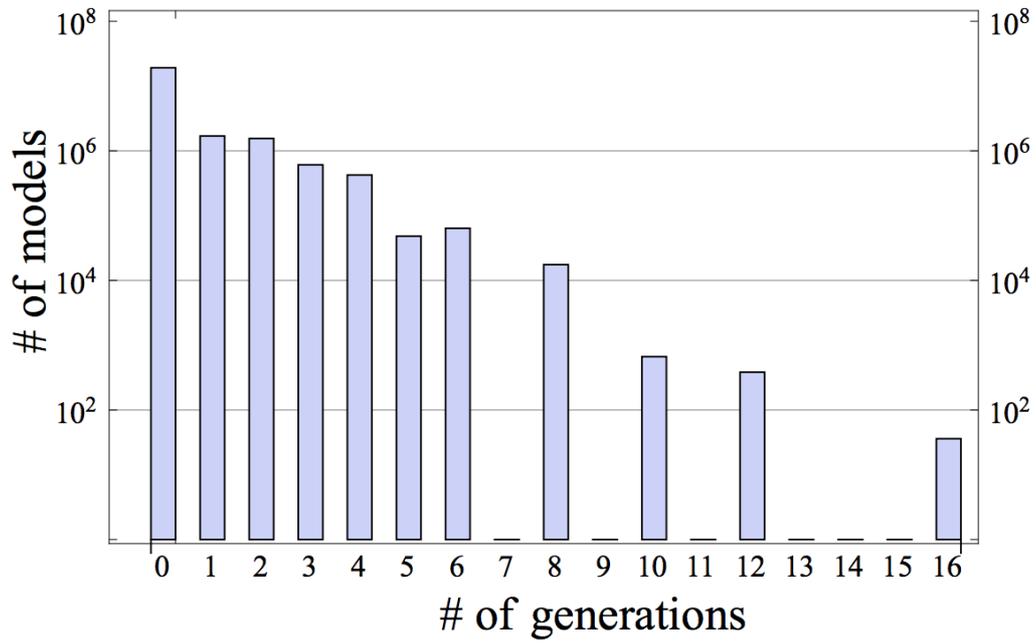}
	\caption{\emph{Number of exophobic models versus number of generations in a random sample of $10^{11}$ Pati-Salam configurations as given in Figure 3 in \cite{acfkr}.}}
	\label{figure2}
\end{figure}

\begin{figure}[!]
	\centering
	\includegraphics[width=140mm]{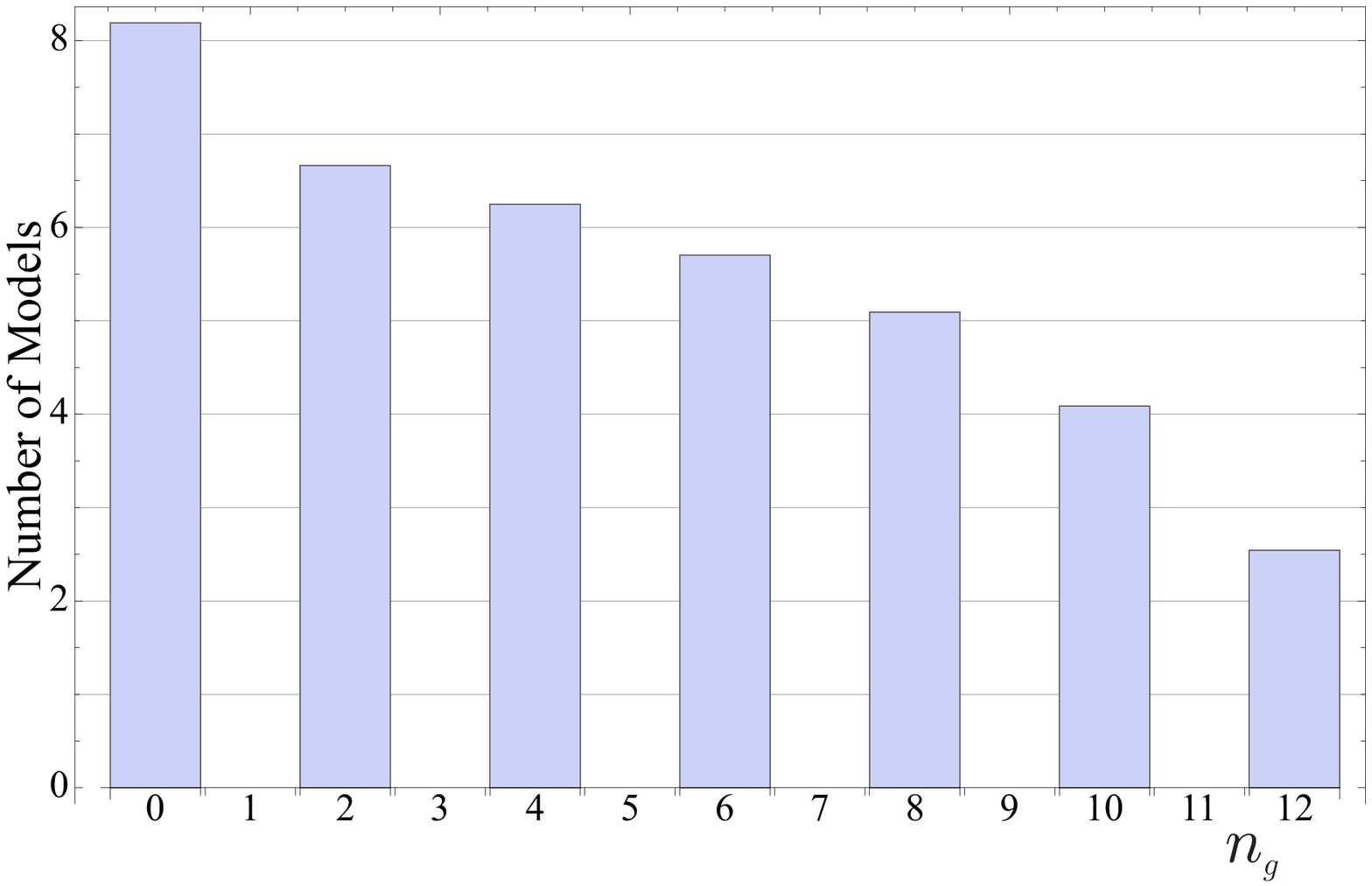}
	\caption{\emph{Logarithm of number of exophobic models versus number of generations in a random sample of $10^{12}$ $SU(5) \times U(1)$ configurations.}}
	\label{figure3}
\end{figure}

\begin{figure}[!]
	\centering
\includegraphics[width=140mm]{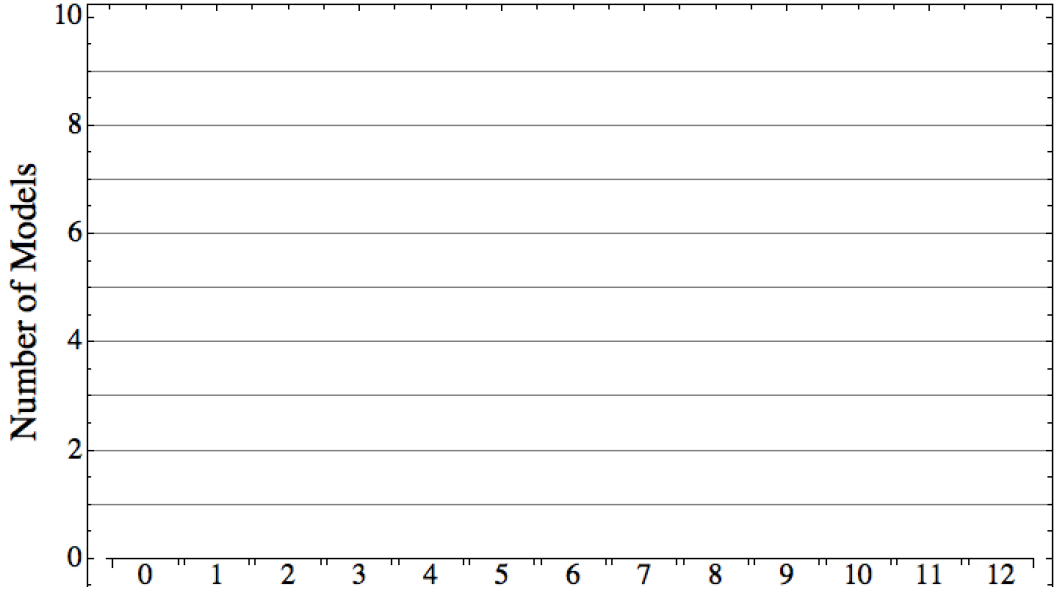}
	\caption{\emph{Number of models versus number of generations in the $SU(4) \times SU(2) \times U(1)$ vacua.}}
	\label{figure4}
\end{figure}

An enormous amount of work has been accomplished over the last decade, for the classification of the symmetric $\mathbb{Z}_2 \times \mathbb{Z}_2$ free fermionic orbifold models \cite{fkr, cfkr, rizos, acfkr, frs, fs}. This was achieved with the aid of computer programs that incorporated the classification methodology as discussed in section 3. In this section, the results of the discrete properties in the landscape of the free fermionic heterotic-string vacua are presented. Originally, the $SO(10)$ models were a success as it showed to have an abundance of 3 generation models, which is given in Figure \ref{figure1} when the classification was done for a random sample of $10^{11}$ string vacua, where discrete properties started to emerge, such as the vanishing of the odd generations above 5 generations and even generations incrementing by 4 integers above 12 generations. After the $SO(10)$ models became a success, the Pati-Salam models were investigated, where the classification results are given in Figure \ref{figure2}. Here the Pati-Salam models contained identical discrete properties as the $SO(10)$ models with the inclusion of all the 24 generations being projected out. This was the case as the spinorial $\bf{16}$ representation of the $SO(10)$ was broken to the $\bf{4}$ and $\overline{\bf{4}}$ representations transforming under the $SU(4)$ of the Pati-Salam, therefore two states were required instead of one to complete a family of generation. Furthermore, equation (\ref{matrix}) required 48 solutions (states) that was not possible due to modular invariance. However, the Pati-Salam models contained many 3 generations that were exophobic just like the $SO(10)$ models. Next in-line was the Flipped $SU(5)$ models which contrary to the $SO(10)$ and Pati-Salam classifications consisting of exophobic 3 generation models, contained only 3 generation models with fractionally charged exotic states in $10^{12}$ scanned string vacua \footnote{Note: although this does not prove that in the free fermionic Flipped $SU(5)$ vacua, exophobic models do not exist, they do not exist in the space of $10^{12}$ vacua explored with the given basis vector in equation (\ref{su5basis}).}, which is given in Figure \ref{figure3}. Another interesting property of the Flipped $SU(5)$ models is that it was more constrained and that the generations followed a logarithmic distribution as well as all the odd generations being projected out. The classification of the $SU(4) \times SU(2) \times U(1)$ models as expected from the two $SO(10)$ breaking basis vectors as in equation (\ref{su421basis}) were even more constrained than the Flipped $SU(5)$ models and in fact forbid 3 generations, which is given in Figure \ref{figure4}. Moreover, this was a rare occurrence in the free fermionic classifications, as the second $SO(10)$ breaking basis vector was unique and that the GGSO projection on the $\bf{16}$ of $SO(10)$ projected all the right-handed particles out and thus complete family of generations were incomplete. The next stage of classification will be the Standard-Like models \cite{frs2}, which require just like the $SU(4) \times SU(2) \times U(1)$ models two $SO(10)$ breaking basis vectors as given in equation (\ref{su3211basis}). The results of this GUT model is in working progress and preliminary scans show interesting results.

\newpage

\subsection{\bf{3 Generation Flipped SU(5) String Vacua}}
\begin{figure}[!]
	\centering
	\includegraphics[width=140mm]{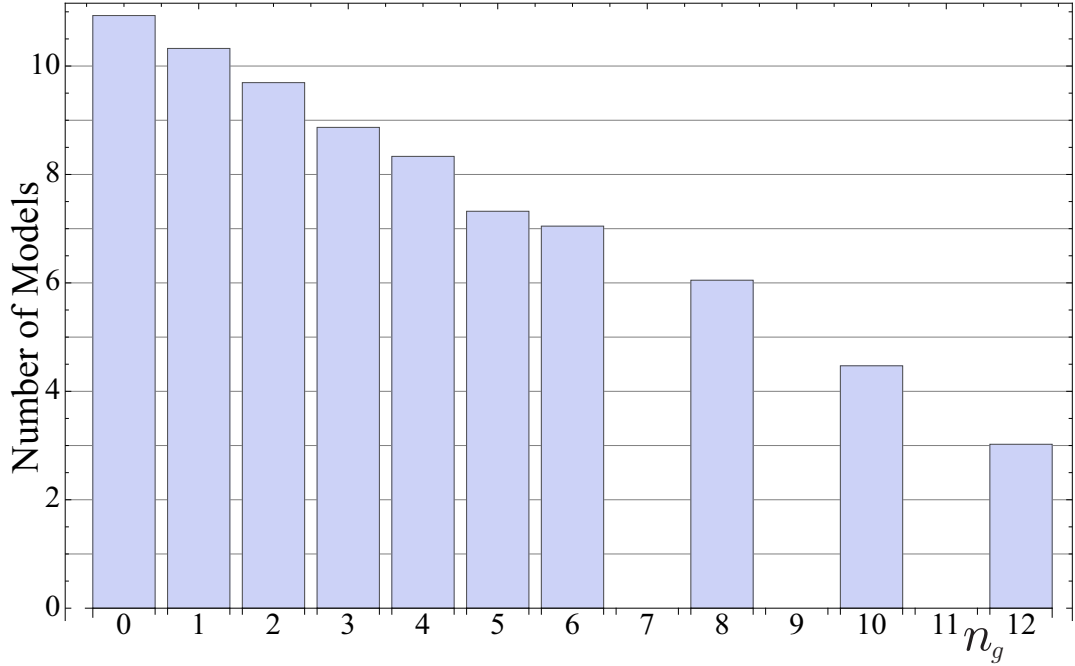}
	\caption{\emph{Logarithm of number of exophilic models versus number of generations    in a random sample of $10^{12}$ $SU(5) \times U(1)$ configurations.}}
	\label{figure5}
\end{figure}

The classification of the Flipped $SU(5)$ vacua that were classified, as shown in Figure \ref{figure3}, resulted in no exophobic 3 generation models. Although this discrete property was remarkable by the non-existence of exophobic odd generations. The Flipped $SU(5)$ models did infact have 3 generation models but with the addition of fractionally charged states in the string spectrum. The distribution again was logarithmic, which is given in Figure \ref{figure5}. As an illustrative example, the following set of GGSO coefficients represent a model with viable phenomenological properties:

\begin{equation} \label{BigMatrix}  (v_i|v_j)\ \ =\ \ \bordermatrix{
	& S&e_1&e_2&e_3&e_4&e_5&e_6&b_1&b_2&z_1&z_2&\alpha\cr
	S  		& 1& 1& 1& 1& 1& 1& 1& 1& 1& 1& 1& 1\cr
	e_1		& 1& 0& 0& 1& 1& 1& 0& 0& 0& 1& 1& 1\cr
	e_2		& 1& 0& 0& 0& 0& 1& 0& 1& 0& 1& 1& 1\cr
	e_3		& 1& 1& 0& 0& 0& 0& 0& 0& 0& 0& 1& 0\cr
	e_4		& 1& 1& 0& 0& 0& 0& 0& 0& 0& 0& 1& 1\cr
	e_5		& 1& 1& 1& 0& 0& 0& 1& 0& 1& 1& 1& 1\cr
	e_6		& 1& 0& 0& 0& 0& 1& 0& 1& 1& 0& 1& 1\cr
	b_1		& 0& 0& 1& 0& 0& 0& 1& 0& 0& 1& 1& 1/2\cr
	b_2		& 0& 0& 0& 0& 0& 1& 1& 0& 1& 0& 0& 1/2\cr
	z_1		& 1& 1& 1& 0& 0& 1& 0& 1& 0& 1& 1& 1\cr
	z_2		& 1& 1& 1& 1& 1& 1& 1& 1& 0& 1& 0& -1/2\cr
	\alpha	& 1& 1& 1& 0& 1& 1& 1& 0& 0& 0& 0& 1\cr
},
\end{equation}
where $[v_i|v_j] = e^{i\pi (v_i|v_j)}$. This model above produces 3 chiral generations with the required heavy and light Higgs breaking states. However, it contains four states that are fractionally charged under the observable $SU(5)$ symmetry. In the $10^{12}$ string vacua classified, the minimum number of fractionally charged states that were given with 3 generations was 4 exotic states, which are unavoidable.

{
	\footnotesize
	\begin{table}{!}
		\begin{tabular}{|c|l|r|c|c|r|}
			\hline
			&Constraints & \parbox[c]{2.5cm}{Total models in sample}& Probability
			&\parbox[c]{3cm}{ Estimated number of models in class}\\
			\hline
			& No Constraints & $1000000000000$ & $1$ &$1.76\times 10^{13}$ \\ \hline
			(1)&{+ No Enhancements} & 762269298719 & $7.62\times 10^{-1}$ & $1.34\times
			10^{13}$ \\  \hline
			(2)&{+ Anomaly Free Flipped $SU(5)$} & 139544182312 & $1.40\times 10^{-1}$ & $2.45\times
			10^{12}$ \\  \hline
			(3)&{+ 3 Generations} & 738045321 & $7.38 \times 10^{-4}$ & $1.30\times
			10^{10}$ \\  \hline
			(4a)&{+ SM Light Higgs} & 706396035 & $7.06 \times 10^{-4}$ & $1.24\times
			10^{10}$ \\\hline
			(4b)&{+ Flipped $SU(5)$ Heavy Higgs} & 46470138 & $4.65 \times 10^{-5}$ & $8.18\times
			10^{8}$ \\\hline
			(5)&{+ SM Light Higgs}& 43624911 & $4.36 \times 10^{-5}$ & $7.67\times
			10^8$ \\ &{+ \& Heavy Higgs}&&&  \\\hline
			(6a)&{\parbox[c]{4cm}{+ Minimal Flipped $SU(5)$ Heavy Higgs}} & 42310396 & $4.23 \times 10^{-5}$ & $7.44\times
			10^8$ \\\hline
			(6b)&{+ Minimal SM Light Higgs } & 25333216 & $2.53 \times 10^{-5}$ & $4.46\times
			10^8$ \\\hline
			(7)&\parbox[c]{4cm}{+ Minimal Flipped $SU(5)$ Heavy Higgs} & 24636896 & $2.46 \times 10^{-5}$ & $4.33\times
			10^8$\\ &{+ \& Minimal SM Light Higgs}&&&  \\\hline
			(8)&{+ Minimal Exotic States} & 1218684 & $1.22 \times 10^{-6}$ & $2.14\times
			10^7$ \\\hline
		\end{tabular}
		\caption{\label{summary} \emph{Phenomenological constraints with respect to Flipped $SU(5)$ models}}
	\end{table}	
}

In table \ref{summary}, detailed statistics are given for the Flipped $SU(5)$ classification, where a sequence of phenomenological constraints are imposed for a total of $10^{12}$ string vacua. Firstly, no enhancements constraint were imposed. This has led to approximately 76.2\% of the models to remain as satisfying this criteria. Next we imposed that the Flipped $SU(5)$ models are anomaly free and then for it to contain 3 chiral generations, this reduced the total viable models by three orders of magnitude. After imposing heavy and light Higgs states to break the Flipped $SU(5)$ GUT gauge group to the Standard Model gauge group and the electroweak breaking respectively, we reduce the order of magnitude by one. Finally, imposing that the minimal number of massless exotics states is 4, we conclude that there are approximately 1 in 1.2 million quasi-string models in the Flipped $SU(5)$ models. To avoid the massless exotics states in the string vacua with 3 generation models, other $SO(10)$ breaking Flipped $SU(5)$ basis vectors are under investigation and work is in progress \footnote{See discussions in \cite{frs}}.

\section{\bf{Conclusion}}
Currently the unification of gravity and the gauge interactions are heavily motivated by string derived models and theories, which continues to provide a viable contemporary framework. For this reason three generations models need to be obtained for phenomenological purposes. A detailed example is still work in progress and might take a long time to find but string theory provides a sea of established quasi-examples that are explored as toy models for achieving a theory of everything. 

The four dimensional free fermionic construction \cite{fff} corresponding to ${\mathbb{Z}}_2 \times {\mathbb{Z}}_2$ orbifold compactifications at special points in the moduli 
space, demonstrated some of the most realistic string models built to date \cite{revamp, alr, slm, lrs}. The classification methodology was originally developed in \cite{gkr} for type II superstrings and then adapted similarly for the free fermionic heterotic-string construction in \cite{fknr,fkr, cfkr, rizos, acfkr, cfr, su62, frs, fs}. This method contains all the possible symmetric ${\mathbb{Z}}_2$-shifts in the internal compactified
directions with the inclusion of six basis vectors $e_i$ for $i=1,...,6$. These six basis vectors enabled the scanning of large number of string vacua. The first classification in \cite{fknr}
was scanned in regards to the chiral $\bf{16}$ and $\overline{\bf{16}}$ spinorial representations 
of the $SO(10)$ GUT group, with the $\bf{10}$ vectorial representation of the $SO(10)$ symmetry included in \cite{fkr}. The spinor-vector duality \cite{fkr, cfkr} discrete symmetry was discovered with the incorporation of the $x$-map \cite{xmap}. The classification methodology itself is based on the writing of the GGSO projections in algebraic forms that elegantly extract the full massless spectrum for any given set of GGSO coefficient configuration. The next series of classifications involved the: Pati-Salam subgroup \cite{acfkr} that led to the discovery of exophobic string vacua \cite{acfkr,cfr}; Flipped $SU(5)$ subgroup \cite{frs} that projected out all the odd generations; $SU(4) \times SU(2) \times U(1)$ subgroup \cite{fs} that projected out all the right-handed particles from the spectrum. The Standard-Like subgroup comes next, which is work in progress \cite{frs2}.

In this paper, I presented the discrete properties that emerged in the landscape of the free fermionic heterotic-string classifications. It was shown that $SO(10)$ models were a great success and the breaking of the $SO(10)$ produced fractionally charged states such as in the Pati-Salam models. Although this reduced the total number of 3 generation models as compared to the $SO(10)$ models, viable models still vastly existed as given in \cite{cfr}. Then adding 1/2 boundary condition assignments to the complex fermions: $\overline{\psi}^{1,...,5}$ complicated the models vastly, as it was shown that the distribution of the number of generation become logarithmic, thus reducing the phenomenologically viable models enormously. Alternative $SO(10)$ breaking basis vectors are being investigated for the Flipped $SU(5)$ subgroups, see \cite{frs} for discussions. The $SU(4) \times SU(2) \times U(1)$ models also consisted of 1/2 boundary condition assignments for the $SO(10)$ breaking that projected out all the right-handed particles. However, this was a unique example as the basis vectors themselves were unique and no alteration can be made to these models. The Standard-Like models are also interesting as it consists of both the Pati-Salam and Flipped $SU(5)$ breaking basis vectors, which is work in progress.

\section{Acknowledgements}
I would like to thank the organisers of the Discrete 2014 conference for their kind invitation. I also would like to thank Alon E. Faraggi and John Rizos for their collaborations and discussions. This work is supported by the STFC studentship award.

\end{document}